\documentclass[prb,twocolumn,nofootinbib,amsmath,amssymb,longbibliography]{revtex4-1}
\usepackage{graphicx}
\usepackage{float}
\usepackage[usenames,dvipsnames]{xcolor}
\begin{document}

\title{
Sliding Dynamics For Bubble Phases on Periodic Modulated Substrates}
\author{
C. Reichhardt and C. J. O. Reichhardt 
} 
\affiliation{
Theoretical Division and Center for Nonlinear Studies,
Los Alamos National Laboratory, Los Alamos, New Mexico 87545, USA
}

\date{\today}

\begin{abstract}
We analyze a bubble forming system composed of particles with
competing long range repulsive and short range attractive interactions
driven over a quasi-one-dimensional periodic substrate.
We find various pinned and sliding phases as a function of substrate
strength and drive amplitude.
When the substrate is weak, a pinned bubble phase appears that
depins elastically into a sliding bubble lattice.
For stronger substrates, we find anisotropic bubbles, disordered bubbles,
and stripe phases.
Plastic depinning occurs 
via the hopping of individual particles from
one bubble to the next in a pinned bubble lattice,
and as the drive increases, there is a transition to a
state where all of the bubbles are moving but are continuously
shedding and absorbing individual particles. This is followed at high
drives by a moving bubble lattice in which the particles can no longer escape
their individual bubbles.
The transition between the plastic and
elastic sliding phases can be detected via signatures in the velocity-force
curves,
differential conductivity, and noise.
When the bubbles shrink due to an increase in the attractive interaction
term,
they fit better inside the pinning troughs and become more strongly pinned,
leading to a reentrant pinning phase.
For weaker attractive terms, the size of the bubbles becomes
greater than the width of the pinning troughs
and the depinning becomes elastic with
a reduced depinning threshold.
\end{abstract}

\maketitle

\section{Introduction}

There are a variety of particle systems that can form bubble phases
exhibiting two length scales,
with the shorter scale determined by the average spacing
between particles that are confined inside an individual bubble,
and the larger scale arising from the assembly of the bubbles themselves
into an ordered lattice.
Bubble lattices typically appear when there is a competition
between attractive and repulsive interactions,
such as short-range attraction and longer-range repulsion
\cite{Mossa04,Reichhardt03,Reichhardt04,Sciortino04,Nelissen05,Liu08,Reichhardt10,CostaCampos13,Zhao13}.
The bubbles can
distort and break up into smaller bubbles as a function of temperature or
as a result of interacting with
quenched disorder \cite{Reichhardt03}.
Bubble phases known as mesophases
arise in colloidal systems that have multiple length scales,
such as multi-step repulsive interactions
\cite{Jagla98,Malescio03,Imperio06,Glaser07}.
Bubble phases can also appear
in superconducting systems for vortices with competing interactions
\cite{Babaev05,Moshchalkov09,Chaves11,Xu11,Sellin13,Komendova13,Meng17},
as well as for magnetic skyrmion-superconducting vortex
hybrids \cite{Neto22}.
Bubbles containing two or more particles can form
when charge ordering occurs
in two-dimensional election systems
\cite{Fogler96, Moessner96,Cooper99,Fradkin99,Friess18,Shingla23}.
In a system with competing interactions,
the bubbles are often only one
of several types of possible
phases, including crystals, stripes, and void lattices,
that arise as a function of particle density or
the ratio of the repulsive and attractive interaction terms
\cite{Seul95,Stoycheva00,Malescio03,Camp03,Glaser07,Reichhardt10}.
The bubble phases occur for lower densities or stronger
attractive interaction terms.
There are also a variety of bubble-like systems where the bubble textures can distort or the shape can change, including emulsions and magnetic textures
such as skyrmions \cite{Jiang15}.

Bubble lattices can interact with a substrate,
which for soft matter systems could be
created using optical trap arrays
\cite{Dholakia02,Reichhardt02a,Brunner02,Gopinathan04,Mikhael10,Juniper16} or
patterned surfaces \cite{Stoop20}.
In solid state systems, ordered substrates can be made using various nanostructuring techniques.
One of the simplest substrate geometries
is a periodic quasi-one-dimensional (q1D)
arrangement of troughs,
which can induce the formation of various crystalline, smectic, and disordered
phases for purely repulsive colloidal particles as a function of
the trough strength and
the ratio of the number of particles to the number of troughs 
\cite{Chakrabarti95,Wei98,Bechinger01,Baumgartl04,Reichhardt05a,Burzle07,Tierno08,Tierno12a}.
Different kinds of vortex patterns and depinning phenomena have been studied in superconducting systems for vortices interacting with periodic q1D
substrates \cite{Dobrovolskiy12,Martinoli78,LeThien16}.
Far less is known about what happens when
bubble or pattern-forming systems with competing interactions
are coupled to a periodic substrate,
and the depinning or sliding dynamics
under a drive have received even less study.
McDermott {\it et al.} \cite{McDermott14} considered a pattern-forming system
with completing repulsion and attraction on periodic q1D
substrates, and found that for a parameter regime
where the system forms a stripe phase in the absence of the
substrate,
several stripe and modulated stripe phases as well as kinks appear
as a function of substrate strength or substrate lattice spacing. 

Here, we examine the statics and dynamics of particles with
competing long-range repulsion and short-range attraction
interacting with a periodic q1D substrate in the limit where
the system forms a bubble phase in the absence of a substrate. For
weak substrates, a triangular bubble lattice appears that
becomes increasingly anisotropic as the substrate strength increases.
When an external drive is applied to the system,
the depinning transition can be elastic, where each bubble retains its
original set of particles, or plastic, where the bubble
lattice remains pinned but individual particles hop from one bubble to
the next.
At higher drives, a bubble shedding phase can appear in which
all of the bubbles are moving but a portion
of the particles can break away from one moving bubble and become attached to
a different moving bubble, while a dynamic reordering transition into
a moving elastic bubble lattice occurs for sufficiently large driving
forces.
In the plastic flow regime, we find a phase that we term
a sliding bubble track phase,
in which stripes of particles remain pinned to the
substrate and form tracks along which the bubbles travel.
We show that the nature of the depinning transition and the net
velocity
depend strongly on whether an individual bubble can fit
into a single substrate trough, and demonstrate that
smaller bubbles are more easily pinned.
This leads to the emergence of 
reentrant pinning as a function of
bubble size, where for constant applied drive the bubbles repin as
they become smaller.
Bubbles that are larger than the width of the
substrate troughs can slide easily in an elastic phase.
Our results should be relevant to a variety of
bubble-forming systems on
q1D substrates,
including electron bubbles, colloidal particles, and
magnetic skyrmions.

\section{Simulation}

We examine a two-dimensional (2D) system
of $N$ particles whose pairwise interactions
have a long-range repulsive term and a short-range attractive term.
The sample is of size $L \times L$ with $L=36$ and has
periodic boundary conditions in the $x$ and $y$
directions.
The particle density is
$\rho = N/L^2$.
The particles interact with a periodic q1D substrate and are
subjected to a dc driving force.
The following overdamped equation governs the dynamics
of particle $i$:
\begin{equation}
\eta \frac{d {\bf R}_{i}}{dt} =
-\sum^{N}_{j \neq i} \nabla V(R_{ij}) + {\bf F}^{s}_{i} +
        {\bf F}_{D} ,
\end{equation}
where the damping term $\eta$ is set to
$\eta=1.0$

The particle-particle interaction potential
is given by
\begin{equation}
V(R_{ij}) = \frac{1}{R_{ij}} - B\exp(-\kappa R_{ij}),
\end{equation}
where
$R_{ij}=|{\bf R}_i-{\bf R}_j|$
and
the location of particle $i (j)$ is
${\bf R}_{i (j)}$.
The first term is
a long range Coulomb repulsion
that will favor
formation of a triangular lattice
of particles in the absence of a substrate.
For computational efficiency, we treat the long range Coulomb
interaction with a Lekner summation, as in
previous work \cite{Reichhardt03,McDermott14}.
The second term is a short range attraction
that falls off exponentially.
At very short range,
the repulsive Coulomb interaction becomes dominant again,
which prevents the particles from collapsing onto a point.
In previous work,
it was shown that particles with the interaction potential
in Eq.~(2)
can form crystal, stripe, bubble, and void lattice
states depending on the values of $\rho$,
$B$, and $\kappa$
\cite{Reichhardt03,Reichhardt03a,Reichhardt04,Reichhardt10,McDermott14}
Here we fix $\kappa = 1.0$ and focus on a
particle density of $\rho = 0.44$.
In the absence of substrate,
this system forms a crystal for $B < 2.0$, a stripe state
for $2.0 < B < 2.25$, and bubbles for $B > 2.25$.
Adding a substrate and/or changing the particle density will
modify the values of $B$ for which these phases appear.

The particles interact with a q1D
substrate with $N_p$ minima. The substrate force is given by
\begin{align}
{\bf F}_s^i = F_p \cos(2\pi x_i/a_p)
\end{align}
where $x_i$ is the $x$ position of particle $i$ and
the spacing between substrate minima is
$a_{p} = L/N_p$.
Here we focus on substrates with $N_{p} = 8.0$,
corresponding to $a_p=4.5$.

We obtain the initial particle configuration by performing simulated
annealing, where the particles are placed in an initial lattice,
subjected to a high temperature, and then slowly cooled. The thermal
forces are represented by Langevin kicks, and after the simulated annealing
is complete, the temperature is set to zero.
After the system has been initialized,
we apply
a driving force of ${\bf F}_{D}=F_D {\bf \hat{x}}$
to all of the particles
and measure
the time-averaged particle velocity in the
direction of drive,
$\langle V\rangle = \sum^{N}_i{\bf v}_i\cdot {\hat {\bf x}}$.
We typically wait
$10^5$ or more time steps until the system has reached a
steady state before taking data,
and we average the velocity over a similar time frame.
Due to the long range interactions, the system can exhibit
transient dynamics over relatively long time scales.

\begin{figure}
\includegraphics[width=\columnwidth]{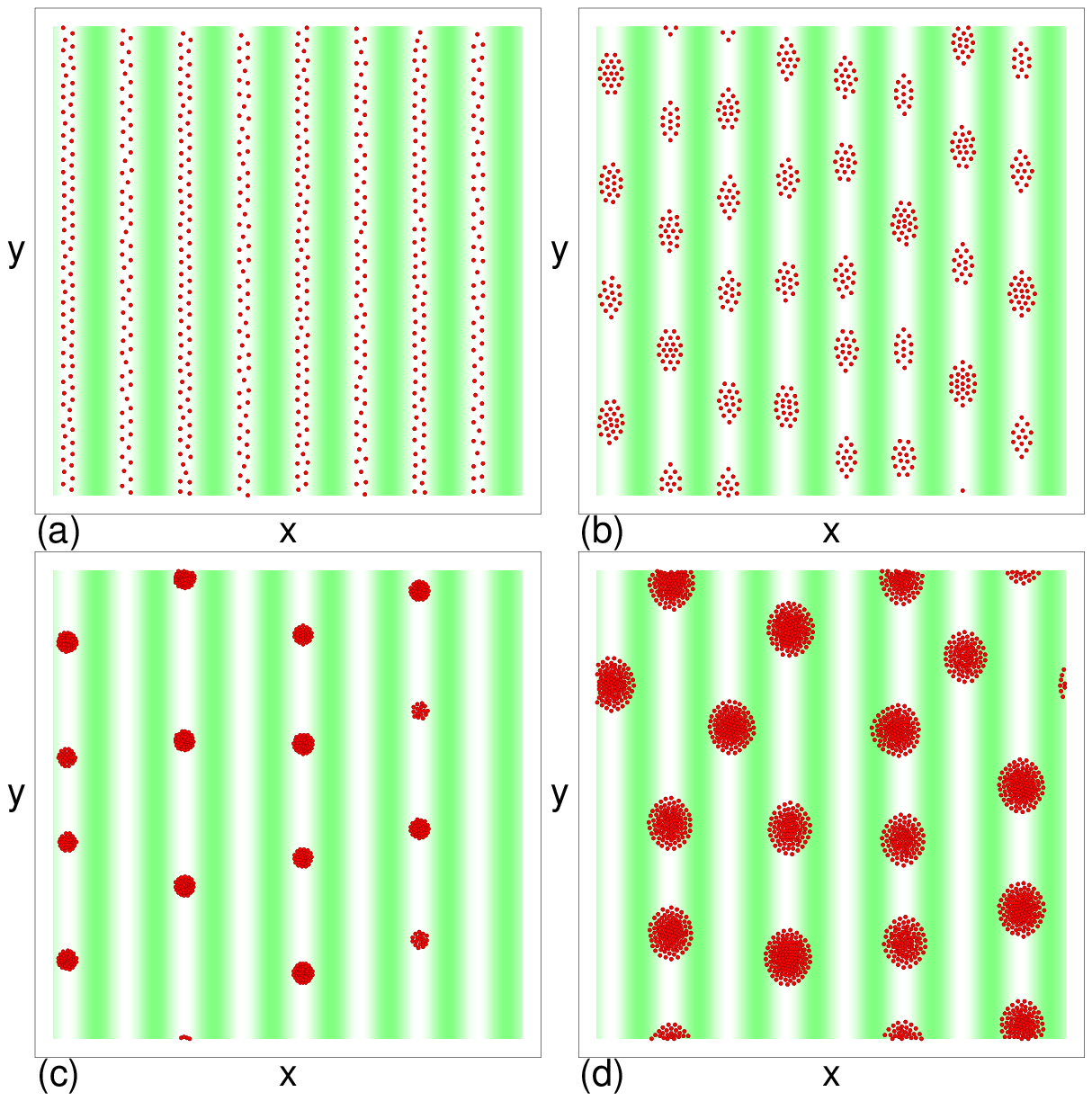}
\caption{The particle positions (red circles) and
the q1D substrate potential (green shading)
for a bubble-forming system
at a particle density of $\rho = 0.44$
with a substrate lattice constant of $a_{p} = 4.5$
for
$F_{p} = 2.0$ and
$F_{D} = 0.0$.
(a) A stripe state at $B = 2.25$.
(b) A bubble phase at $B = 2.85$.
(c) More compact bubbles at $B = 6.0$.
(d) The $B=2.85$ sample from (b) at
$\rho =  1.38$, where larger bubbles appear.
} 
\label{fig:1}
\end{figure}

\section{Results}
In Fig.~\ref{fig:1}(a-c),
we illustrate the particle positions and substrate potential
for a system with
$F_{p} = 2.0$, $\rho = 0.44$,
and $a_{p} = 4.5$ at $F_{D} = 0.0$.
For $B=2.25$ in Fig.~\ref{fig:1}(a),
the particles form stripes that are
aligned with the substrate.
In Fig.~\ref{fig:1}(b) at $B = 2.85$, bubbles appear that have 
an anisotropic  shape due to the confinement from the substrate.
At $B=6.0$ in Fig.~\ref{fig:1}(c),
the bubbles are much smaller and can fit
better into the sinusoidal
substrate minima.
Figure~\ref{fig:1}(d) shows the bubble phase sample
with $B=2.85$ from Fig.~\ref{fig:1}(b)
at a higher particle density of $\rho = 1.38$,
where the bubbles are more compact and
each contain a larger number of particles.

\begin{figure}
\includegraphics[width=\columnwidth]{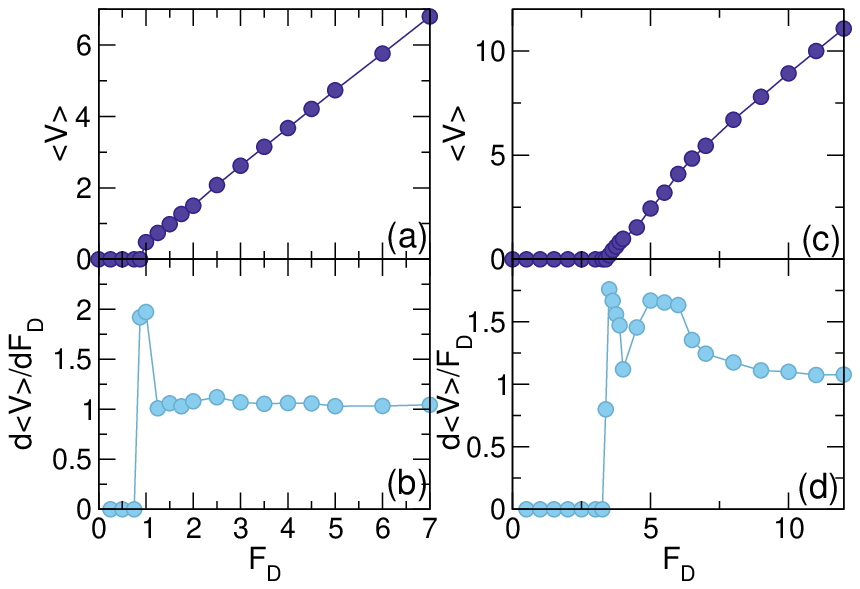}
\caption{(a) The average velocity per particle $\langle V\rangle$
vs $F_{D}$ for the system in Fig.~\ref{fig:1}(b)
with $a_p=4.5$,
$\rho = 0.44$, $B=2.85$, and $F_{p} = 2.0$, where the
depinning is elastic.
(b) The corresponding $d\langle V\rangle/dF_{D}$ vs $F_{D}$ curve.
(c) $\langle V\rangle$ vs $F_D$ for the same system but with
$F_{p} = 5.0$, where the depinning is plastic.
(d) The corresponding $d\langle V\rangle/dF_D$ vs $F_D$
curve indicates that there is  a two-step depinning process.}
\label{fig:2}
\end{figure}

We next focus on the depinning of the bubble phase for $B \geq 2.25$
by measuring the average velocity as a function of external drive
for fixed $\rho = 0.44$ and $B  = 2.85$.
In Fig.~\ref{fig:2}(a), we plot the
velocity-force curve for the system in
Fig.~\ref{fig:1}(b).
Near $F_{D} = 0.85$, 
the bubbles depin elastically, with all of the particles remaining in their
original bubble both during the depinning process and at higher drives.
Figure~\ref{fig:2}(b) shows that the corresponding
$d\langle V\rangle/dF_{D}$ versus $F_D$ curve has
a sharp peak near depinning,
while at higher drives the velocity-force curve becomes linear and
$\langle V\rangle \propto F_{D}$.
In systems that depin elastically, there
is generally only a single peak in the
differential velocity-force curves \cite{Reichhardt17}.
When we increase the substrate strength,
we see a transition to plastic flow where the motion above depinning
consists of individual particles hopping from one pinned
bubble to the next, followed at higher drives by a state
where moving bubbles shed and reabsorb individual particles as they
travel through the system.
In Fig.~\ref{fig:2}(c,d), we plot
the velocity-force and differential velocity-force curves
for the same system
from Fig.~\ref{fig:2}(a) but at $F_p=5.0$,
where the depinning is plastic. Here,
there are two peaks in the differential velocity-force curve,
and the velocity-force curve does not exhibit linear behavior
until $F_{D} > 8.5$.

\begin{figure}
\includegraphics[width=\columnwidth]{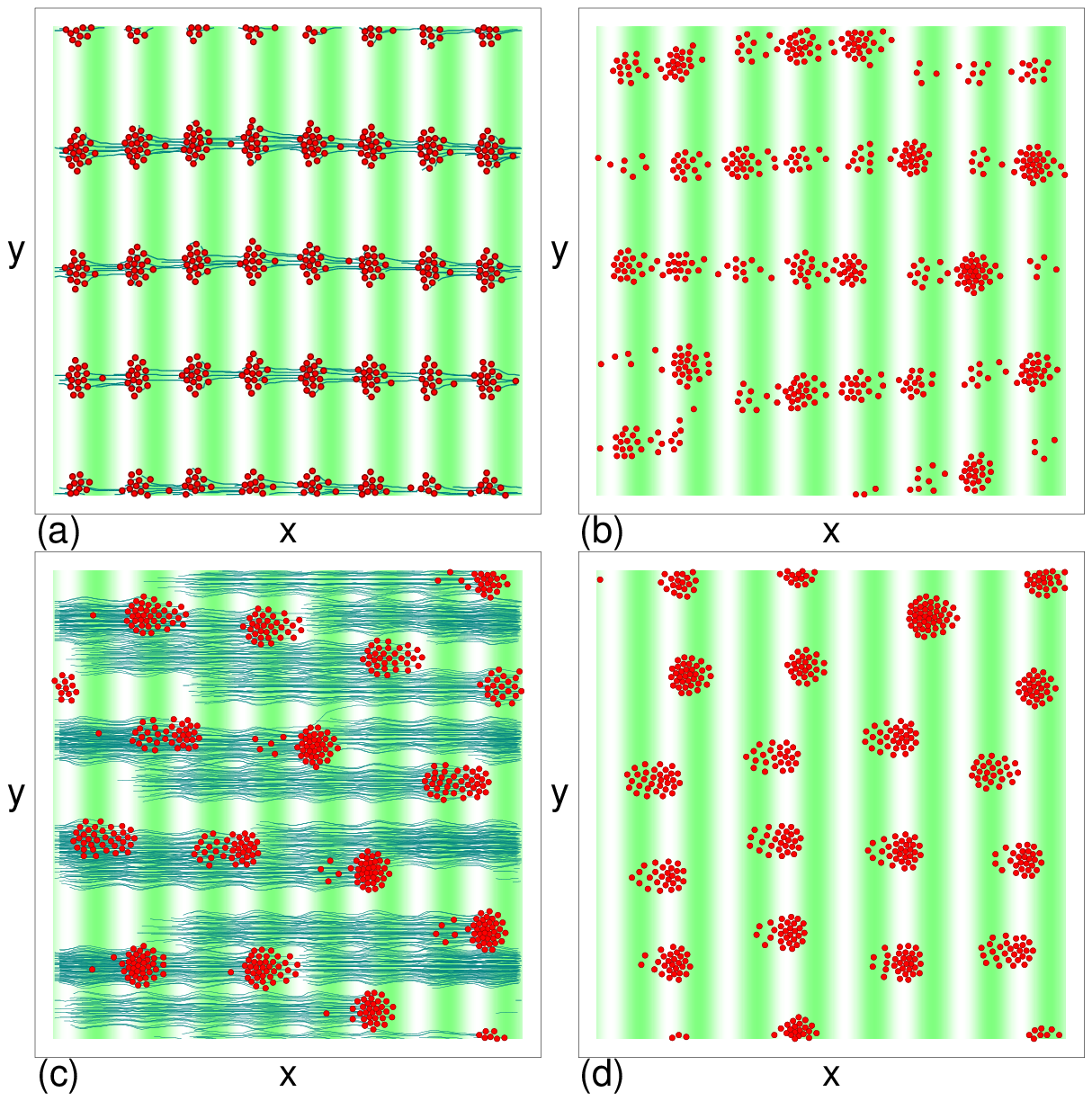}
\caption{The particle positions (red circles), trajectories (lines),
and q1D substrate potential (green shading) for the system
in Fig.~\ref{fig:2}(c,d) with 
$a_p=4.5$, $\rho=0.44$,
$B=2.85$, and
$F_p=5.0$.
(a) At $F_{D} = 3.75$ or
$F_D/F_p=0.75$, the bubbles are pinned,
but individual particles are hopping from one bubble to the next.
(b) At $F_{D}/F_{p} = 1.1$, all
of the particles are moving but the bubble structure is disordered.
For clarity, the trajectory lines are not shown.
(c) At $F_{D}/F_{p} = 1.4$,
organized bubbles reform but can shed and reabsorb
individual particles, which are able to use the shedding mechanism to
jump from one bubble to another.
(d) At $F_{D}/F_{p} = 2.2$, the bubbles are fully formed and there
is no shedding. For clarity, the trajectory lines are not shown.}
\label{fig:3}
\end{figure}

In Fig.~\ref{fig:3}(a), we show the
particle locations and trajectories
for the system in Fig.~\ref{fig:2}(c,d) at $F_{D} = 3.75$,
just above the first peak in the
$d\langle V\rangle/dF_{D}$ curve. The system forms a rectangular
lattice of pinned bubbles, and individual particles are able to hop
from one bubble to the next.
As $F_{D}$ increases, the number of particles participating in this
hopping process increases
until $F_{D}/F_{p} >  1.0$, at which point all of the particles
are able to move at the same time. When this happens,
the bubble structure is partially broken up, as shown in Fig.~\ref{fig:3}(b) at
$F_{D}/F_{p} = 1.1$.
The second peak in the
$d\langle V\rangle/F_{D}$ curve
thus corresponds to the drive at which
all of the particles become able to flow simultaneously.
As $F_{D}$ further increases,
the bubble structure reassembles,
as shown in Fig.~\ref{fig:3}(c) at $F_{D}/F_{p} = 1.4$,
and each bubble can shed individual particles that
proceed to jump from one
moving bubble to another, where the particles are reabsorbed.
In general, the individual particles move more slowly than the bubbles.
For $F_{D} > 1.7$, the system forms a bubble lattice
where the hopping process no longer occurs and all particles remain
trapped within individual bubbles,
as shown in Fig.~\ref{fig:3}(d) at $F_{D}/F_{c} = 2.2$.
The bubble lattice formation occurs at the same drive that marks the
transition to linear behavior of the velocity-force curve.
In the moving bubble phase, the individual bubbles are anisotropic
and are elongated in the driving direction.

\begin{figure}
\includegraphics[width=\columnwidth]{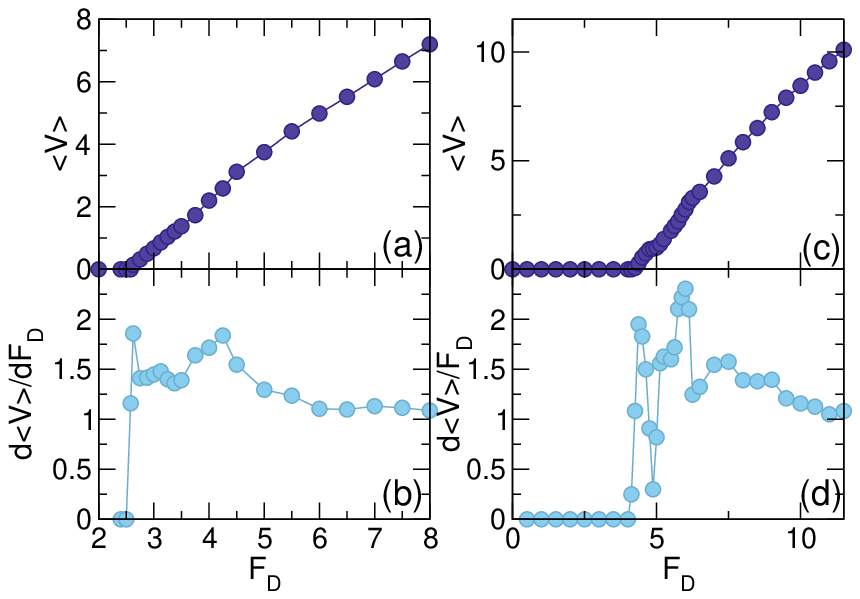}
\caption{(a) $\langle V\rangle$ vs $F_{D}$
for the system from Fig.~\ref{fig:2} with $a_p=4.5$, $\rho=0.44$, and
$B=2.85$ but at $F_p=4.0$.
(b) The corresponding $d\langle V\rangle/dF_D$ vs $F_D$ curve
showing a multiple step depinning process.
(c) $\langle V\rangle$ vs $F_D$ for the
same system but with $F_p=6.0$.
(d) The corresponding $d\langle V\rangle/dF_D$ vs $F_D$ curve.
}
\label{fig:4}
\end{figure}

The nonlinear velocity-force curve and
the double peak feature in the differential
velocity-force curve is a general feature of systems that
exhibit plastic depinning \cite{Reichhardt17}.
In Fig.~\ref{fig:4}(a,b), we plot the velocity-force
and differential velocity-force curves for the system
from Fig.~\ref{fig:2} but at $F_{p} = 4.0$,
while Fig.~\ref{fig:4}(c,d) shows $\langle V\rangle$ versus $F_D$ and
$d\langle V\rangle/F_D$ versus $F_D$ for the same system at
$F_{p} = 6.0$, where the multiple peak feature
in $d\langle V\rangle/F_D$ is even more pronounced.
From the images and the features in the transport curves,
we can identify
four dynamical phases
that appear as a function of applied drive and substrate strength.
These are: a pinned bubble phase where the velocity is zero;
an intra-bubble hopping phase where the bubbles are pinned
but individual particles can hop from one bubble to the next;
a disordered moving partial bubble phase
where all of the particles are moving but the bubble structure has
been disrupted by a shedding and reabsorption process;
and a high drive elastically moving bubble lattice.

\begin{figure}
\includegraphics[width=\columnwidth]{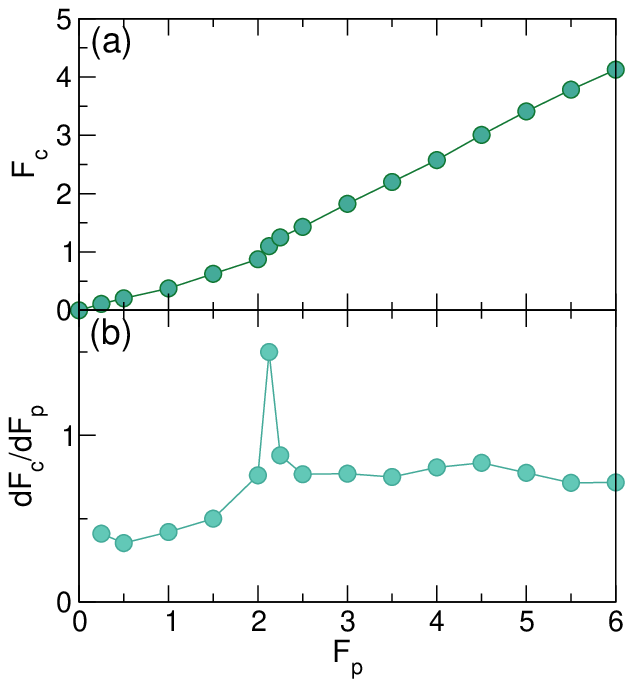}
\caption{(a) The depinning force $F_{c}$ vs $F_{p}$ for the
system in Figs.~\ref{fig:2} and \ref{fig:4}
with $a_p=4.5$, $\rho=0.44$, and $B=2.85$.
(b) The corresponding $dF_{c}/dF_{p}$ vs $F_{p}$
curve. The peak corresponds to the transition
from elastic depinning to plastic depinning with increasing $F_p$. }
\label{fig:5}
\end{figure}

In Fig.~\ref{fig:5}(a), we show the depinning
threshold $F_{c}$ versus pinning strength $F_{p}$ for the
system in Figs~\ref{fig:2} and \ref{fig:4},
while Fig.~\ref{fig:5}(b) shows the  corresponding
$dF_{c}/dF_{p}$ versus $F_{p}$ curve.
The depinning is elastic for $F_{p} < 2.25$ and
plastic for $F_{p} \geq 2.25$.
The elastic-to-plastic 
transition is marked by a jump up in the depinning
force and a peak in $dF_{c}/dF_{p}$.
A sharp increase
in the depinning threshold
at the transition from elastic to plastic behavior has been well studied
in other systems with random or periodic substrates
that exhibit depinning phenomena \cite{Reichhardt17}.
Additionally, for $F_{p} \geq 2.25$, we find that the
depinning threshold
increases linearly according to $F_{c} \approx F_{p}$,
as expected for plastic depinning  \cite{Reichhardt17}.  

\begin{figure}
\includegraphics[width=\columnwidth]{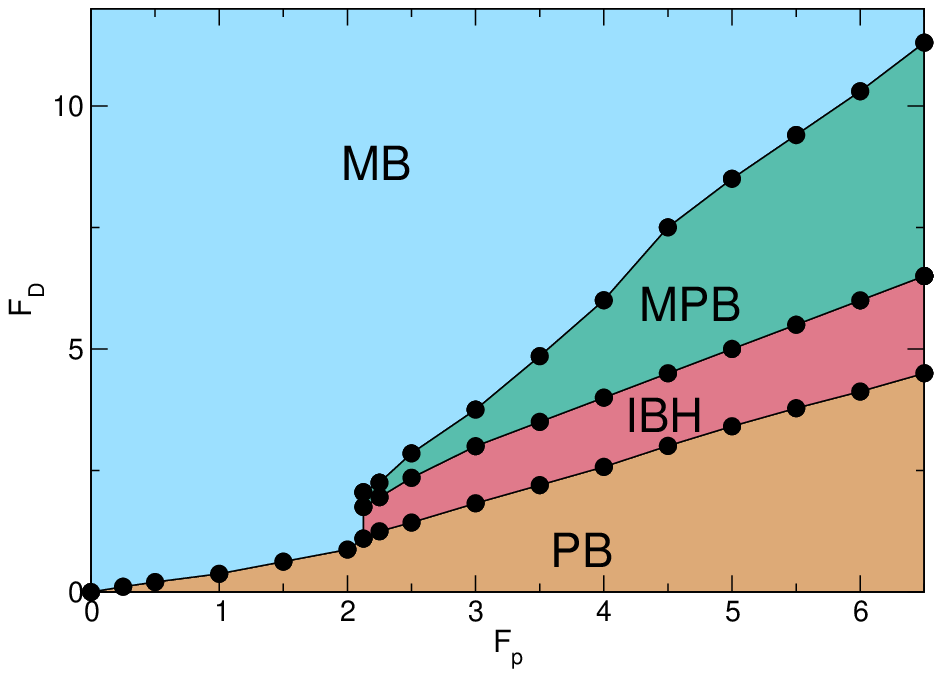}
\caption{Dynamic phase diagram as a function of
$F_D$ vs $F_P$ for the system from Fig.~\ref{fig:4} with
$a_p=4.5$, $\rho=0.44$, and $B=2.85$, where we highlight  
the moving bubble lattice (MB) phase,
an intra-bubble hopping (IBH) phase,
a moving partial bubble (MPB) phase,
and a pinned bubble (PB) phase. }
\label{fig:6}
\end{figure}

From the particle structures and the
features in the transport curves,
we can construct a dynamical phase diagram
highlighting the different dynamical regimes,
as shown in Fig.~\ref{fig:6} as a function of $F_{D}$ versus $F_{p}$
for the system from Figs.~\ref{fig:2} and \ref{fig:4}.
For $F_{p} < 2.25$, there is an elastic depinning transition
directly from a pinned bubble (PB) lattice
to a moving bubble (MB) lattice,
while for $F_{p} \geq 2.25$, an initial depinning transition
takes the system into the intra-bubble hopping (IBH) phase,
a second depinning transition results in the appearance
of the disordered moving partial bubble (MPB) phase,
and at higher drives a dynamic reordering transition
occurs into the moving bubble (MB) lattice.
Dynamical ordering at high drives
in systems with plastic depinning transitions
has been observed in systems with both random
and ordered substrates \cite{Reichhardt17}.

\section{Effect of Bubble Size on Dynamics}

We next consider the effect of changing the strength $B$ of the
attractive interaction term.
For larger $B$, the bubbles shrink in size and become more rigid,
as shown in Fig.~\ref{fig:1}(c).
We note that individual bubbles can capture an increased
number of particles even
as the bubble radius decreases with increasing $B$;
however, because the Coulomb interaction becomes dominant again at
small length scales, the particles are unable to assemble into
a single large bubble but instead form a collection of bubbles.

\begin{figure}
\includegraphics[width=\columnwidth]{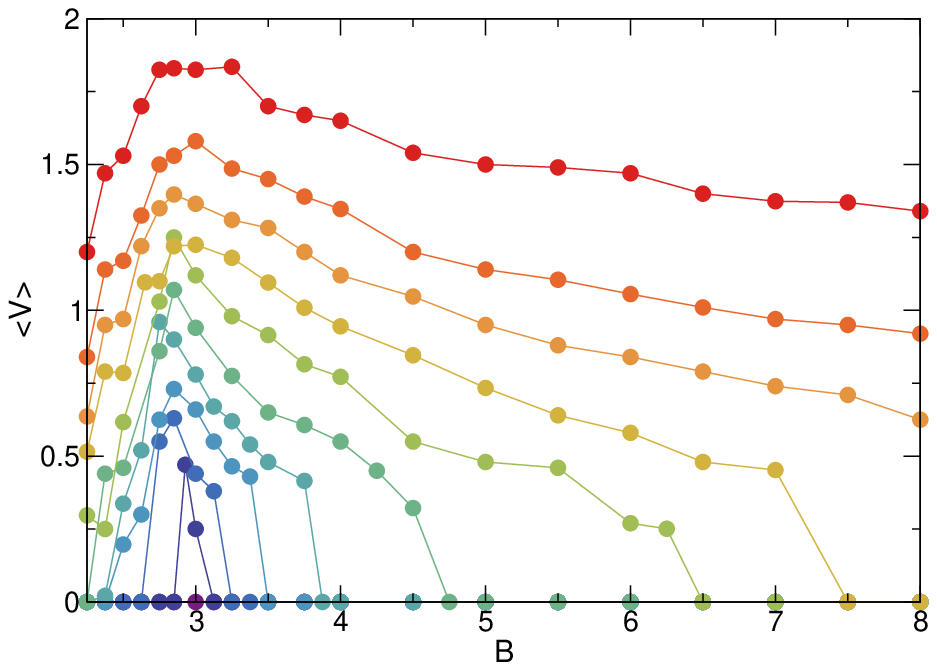}
\caption{$\langle V\rangle$ vs $B$ for a system with
$a_p=4.5$,
$\rho=0.44$, and $F_p=0.1$ at
$F_{D} = 0.875$, 1.0, 1.125, 1.25, 1.375, 1.5, 1.625, 1.75,
1.875, 2.0, and $2.25$, from bottom to top.
Here, $\langle V\rangle$ is nonmonotonic,
and there is a reentrant pinning transition
at higher $B$.}
\label{fig:7}
\end{figure}

We first consider the case for $F_{p} = 2.0$,
where the depinning transition is elastic.
In Fig.~\ref{fig:7}, we plot
$\langle V\rangle$ versus $B$
in a sample with $\rho=0.44$ and $F_p=0.1$ at $F_{D} = 0.875$, 1.0, 1.125,
1.25, 1.375, 1.5, 1.625, 1.75, 1.875, 2.0, and $2.25$.
When $F_{D} <  1.375$, the velocity drops to zero for low values
of $B$.
In this low $B$ pinned state, a
strongly pinned stripe phase appears, 
and the
depinning and sliding dynamics of this
stripe phase will be described in a separate paper.
For each value of $F_D$, $\langle V\rangle$ has a non-monotonic
dependence on $B$, with
a maximum appearing near $B  = 3.0$ followed by a velocity decrease.
When $F_{D} < 1.875$,
at higher $B$ the velocity drops completely to zero, indicating that
the bubbles have become pinned.
This result indicates
that there can be both a low $B$ pinned state and a
high $B$ reentrant pinned
state.

\begin{figure}
\includegraphics[width=\columnwidth]{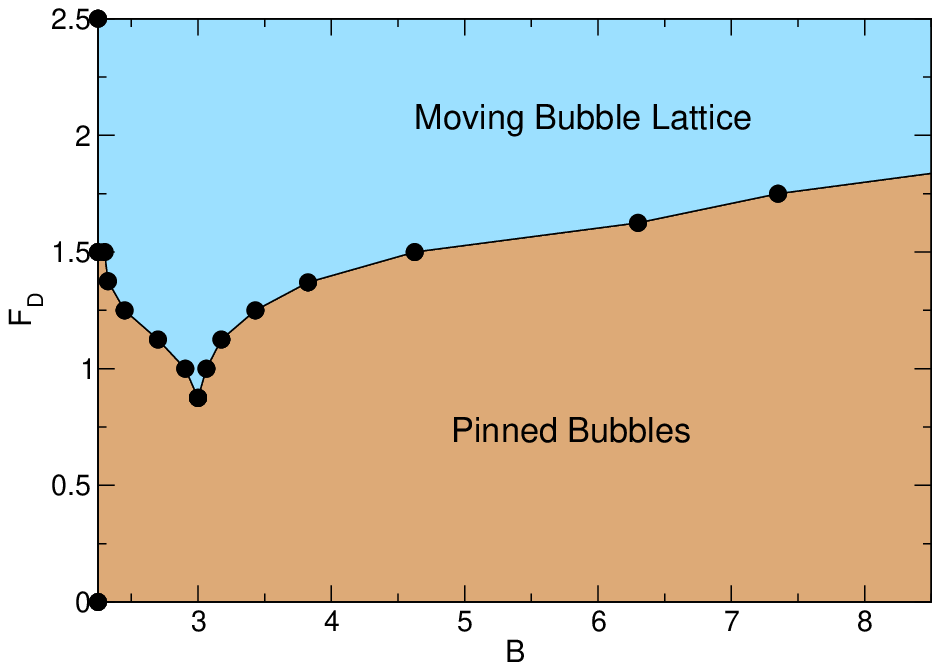}
\caption{Dynamic phase diagram as a function of
$F_{D}$ vs $B$ showing the pinned bubble and moving bubble lattice regimes
for the system in Fig.~\ref{fig:7} with 
$a_p=4.5$, $\rho=0.44$ and $F_p=0.1$.
A dip appears near $B = 3.0$ where the bubbles are less well pinned.
}
\label{fig:8}
\end{figure}

The non-monotonic behavior of the velocity
and the reentrant pinning occur because of two effects: the ability
of the bubble to distort, and the change in size of the bubble.
For $B < 3.0$, the bubbles are large but can partially
distort to fit inside the pinning troughs as elongated or anisotropic
bubbles.
As $B$ increases, the bubbles shrink but also become stiffer,
and are thus less able to accommodate anisotropic distortions.
Near $B = 3.0$, the bubbles become too round to elongate enough
to fit completely into an individual pinning trough,
reducing the effectiveness of the pinning.
For $B> 3.0$, the bubbles remain round, but the
bubble radius decreases, as shown in Fig.~\ref{fig:1}(c).
These smaller bubbles can fit more easily into the 
pinning troughs, increasing the effectiveness of the pinning,
and the system starts to act more like a lattice of point particles.
Thus, the smaller bubbles are better pinned even when the number
of particles in each individual bubble remains the same.
In Fig.~\ref{fig:8}, we plot a dynamic phase diagram
as a function of $F_{D}$ versus $B$
showing the pinned bubbles and moving bubble lattice phases.
A dip in the transition point appears
near $B = 3.0$, where the pinning effectiveness of the substrate
is the most greatly reduced by the shape of the bubbles.
For $B \leq 2.25$, the system forms a moving stripe phase.

\begin{figure}
\includegraphics[width=\columnwidth]{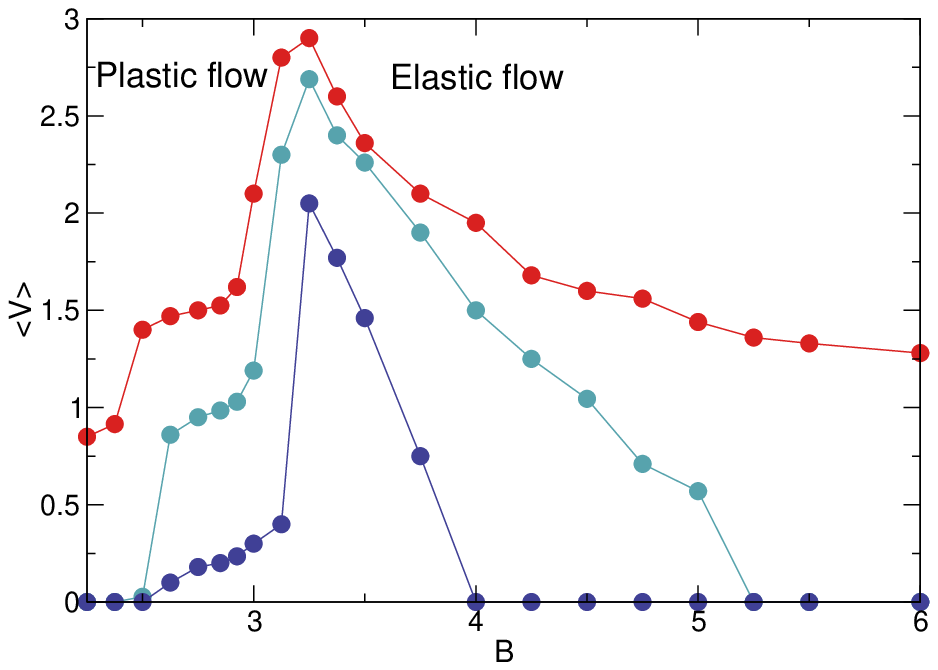}
\caption{$\langle V\rangle$ vs $B$ for
the system from Fig.~\ref{fig:7} with $a_p=4.5$ and $\rho=0.44$
at $F_p=5.0$ for $F_D=3.5$, 4.0, and 5.0, from bottom to top.
The depinning is plastic
for $B<3.25$ and elastic for $B \geq 3.25$,
and there is a reentrant pinning regime at high $B$.}
\label{fig:9}
\end{figure}

\begin{figure}
\includegraphics[width=\columnwidth]{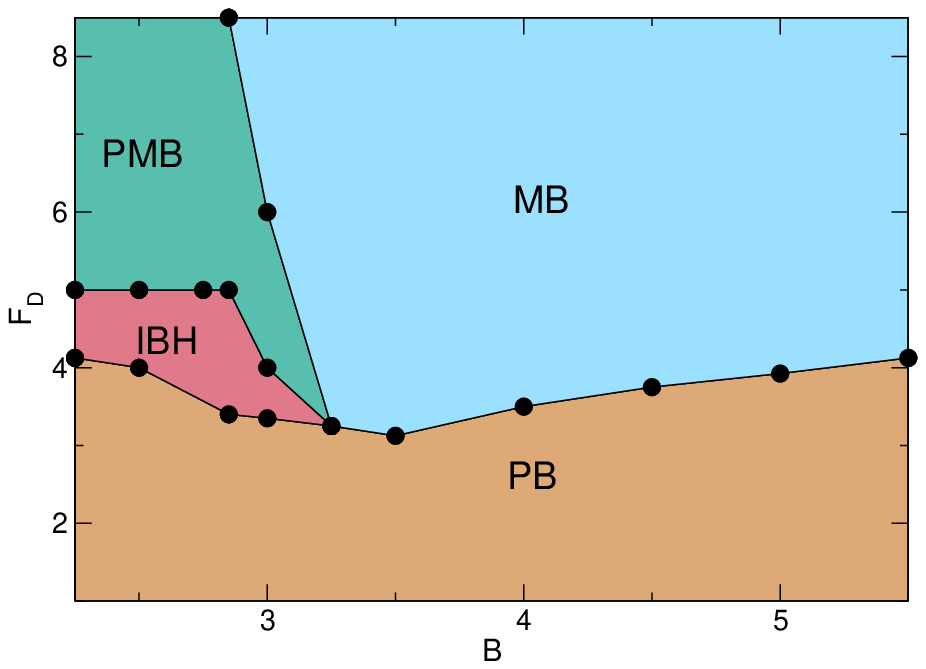}
\caption{Dynamic phase diagram as a function of
$F_D$ vs $B$  
for the system in Fig.~\ref{fig:9} with
$a_p=4.5$, $\rho=0.44$ and $F_p=5.0$
showing the pinned bubble (PB) phase, intra-bubble hopping (IBH) phase,
moving partial bubble (MPB) phase, and elastic moving bubble (MB) phase.}
\label{fig:10}
\end{figure}

In Fig.~\ref{fig:9} we plot $\langle V\rangle$ versus $B$ for
the same system from Fig.~\ref{fig:7} but with $F_p=5.0$ at
$F_{D} = 3.5$, $4.0$, and $5.0$.
For this system,
the depinning is plastic for $B < 3.25$ and elastic for
$B \geq 3.25$.
The velocities are lower within the plastic or IBH regime
and show a jump up at the entrance to the
elastic regime, with a peak velocity appearing near $B = 3.25$
followed by a velocity drop at larger $B$
as the bubble size decreases. There is
a reentrant pinning regime at high values of $B$.
The smaller bubbles
are less likely to undergo
plastic deformations or permit bubble-to-bubble hopping
because the attractive interaction forces generate
a greater barrier for
individual particles to jump out of a bubble.
From the flow patterns and the features
in the transport curves,
in Fig.~\ref{fig:10} we construct a dynamic phase diagram 
for the system in Fig.~\ref{fig:9} as
a function of $F_D$ versus $B$.
The depinning force is lowest
near $B=3.5$, close to the elastic to plastic transition,
and the drive that must be applied to the plastic phase in order
to dynamically reorder the system increases with decreasing
$B$.
As $B$ increases in the elastic regime,
the depinning threshold increases
because the bubble size is diminishing.

\begin{figure}
\includegraphics[width=\columnwidth]{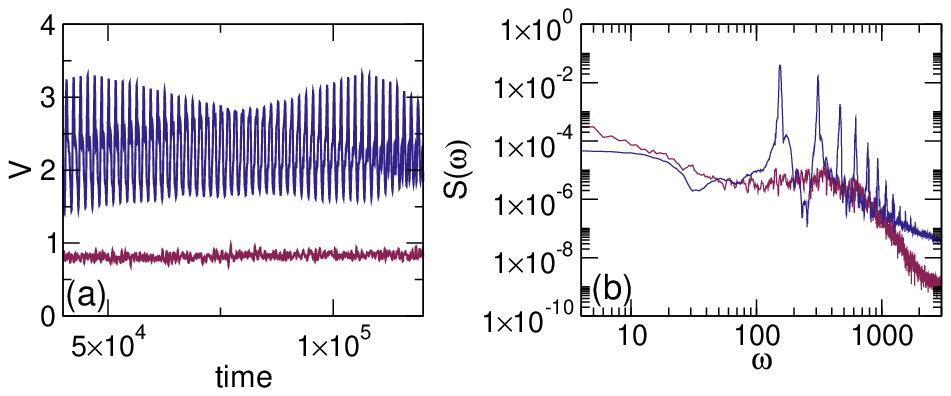}
\caption{Velocity time series $V$ vs time
for the system in Fig.~\ref{fig:10} with
$a_p=4.5$, $\rho=0.44$, $F_p=5.0$,
and $F_D=4.0$
at $B = 2.75$ (red) in the IBH or plastic flow phase
and $B = 3.5$ (blue) in the moving bubble lattice or MB phase.
There is a strong time periodic signal in the MB phase.
(b) The corresponding power spectra $S(\omega)$
shows strong peaks in the moving bubble phase.}
\label{fig:11}
\end{figure}

Another way to characterize the
plastic and elastic flow regimes is to measure the power spectra of the
velocity noise fluctuations from time series data.
In Fig.~\ref{fig:11}(a), we plot the time
dependent velocity for the system in Figs.~\ref{fig:9}
and \ref{fig:10} at $B = 2.75$,
which corresponds to the inter-bubble hopping
plastic flow phase
where the velocity signal has no significant features.
We also plot $V$ as a function of time for
a sample with $B=3.5$ in the moving bubble phase,
where there is a strong periodic
velocity signal consistent with
the washboard signature
expected for an elastic solid moving over a periodic potential.
Figure~\ref{fig:11}(b) shows the power spectra
$S(\omega)=|\sum V(t) e^{-i\omega t}|^2$
for the
two velocity signals.
In the intra-bubble hopping phase, there is $1/f$ noise at lower frequencies,
while a strong narrow band noise signal appears for the moving bubble phase. 
This indicates that moving bubbles should produce a washboard signal,
while the intra-bubble hopping flow is more disordered.
We find similar noise features
for plastic versus elastic flow for other parameters.

\begin{figure}
\includegraphics[width=\columnwidth]{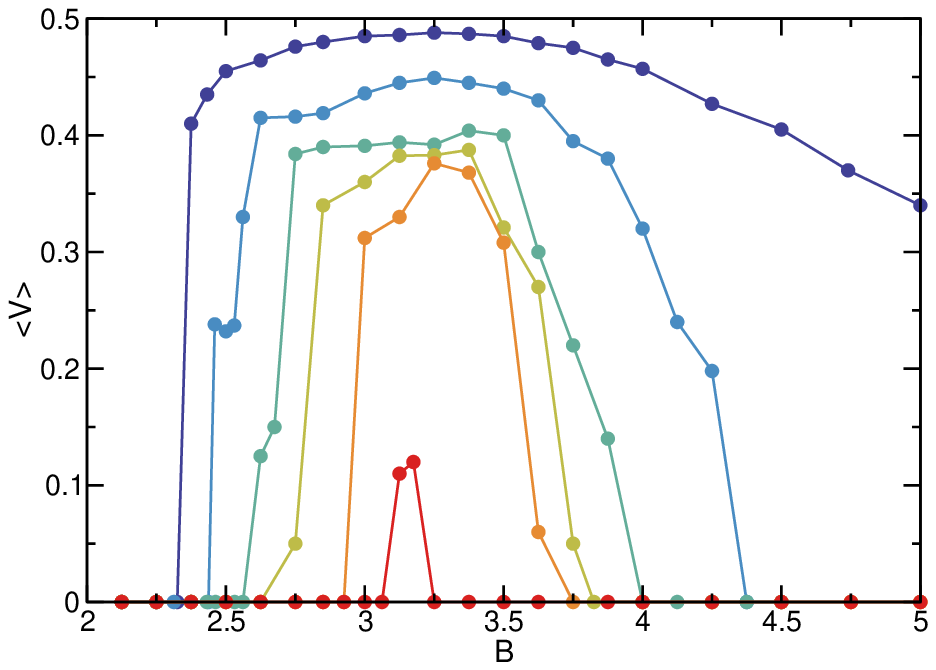}
\caption{$\langle V\rangle$ vs $B$ for a system
with $F_{D} = 0.5$, $\rho = 0.44$,
and a smaller pinning lattice constant of
$a_p = 2.11$
for $F_{p}= 1.0$, 2.0, 3.0, 4.0, 5.0, and $6.0$, from top to bottom.
At low $B$, the system forms a pinned stripe phase as
illustrated in Fig.~\ref{fig:13}(a) for $B = 2.25$ and $F_{p} = 2.0$.
When the velocity becomes finite, the system depins
plastically into the
sliding bubble track phase shown in Fig.~\ref{fig:13}(b)
for $B = 2.5$ and $F_{p} = 2.0$.
At high velocities,
the system forms a bubble phase where
the bubbles are close to twice the size of the pinning lattice constant,
as shown in Fig.~\ref{fig:13}(c) for $B = 2.75$ and $F_{p} = 2.0$.
For higher $B$, the system reenters a pinned state when the bubbles
become small enough to fit inside a single pinning trough,
as shown in Fig.~\ref{fig:13}(d) for $B = 4.5$ and $F_{p} = 2.0$.
}
\label{fig:12}
\end{figure}

\begin{figure}
\includegraphics[width=\columnwidth]{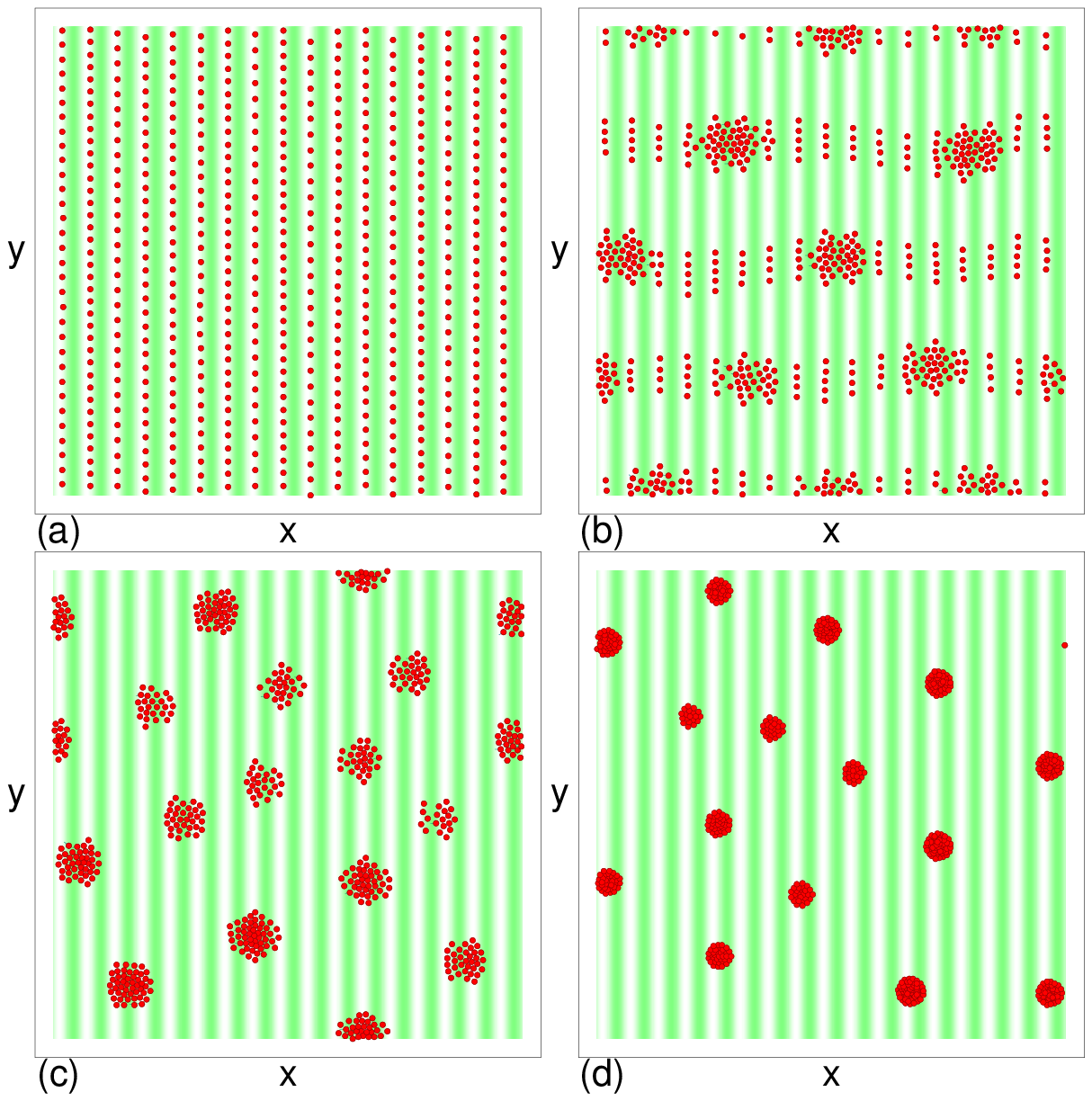}
\caption{The particle positions (red circles) and q1D substrate potential (green
shading) for the system in Fig.~\ref{fig:12} with $a_p=2.11$, $\rho=0.44$,
$F_{p} = 2.0$, and $F_{D} = 0.5$.
(a) A pinned stripe phase at $B = 2.25$.
(b) A plastic sliding bubble track phase at $B = 2.5$.
Pinned particles form stripe tracks and
the bubbles move along these stripe tracks.
(c) A moving bubble phase at $B = 2.75$ where the bubble size
is greater than the width of the substrate troughs.
(d) A pinned bubble lattice at $B = 4.5$ where the bubbles are small
enough to fit within individual pinning troughs.}
\label{fig:13}
\end{figure}

To further explore how the bubble size affects the pinning
effectiveness,
we consider a substrate with a smaller lattice constant
at the same particle density and the same range of $B$ values.
In Fig.~\ref{fig:12} we plot $\langle V\rangle$ versus $B$
for a system with $\rho = 0.44$, fixed $F_{D} = 0.5$, and a
substrate lattice constant of $a_p=2.11$, half as large as what was used
for the results presented up to this point, at substrate strengths of
$F_{p} = 2.0$, 3.0, 4.0, 5.0, and $6.0$.
For $F_{p} = 2.0$, the system is pinned when $B < 2.4375$,
and the particles
form a 1D stripe-like pattern as shown in
Fig.~\ref{fig:13}(a) for $B = 2.25$.
A new plastic flow state, distinct from that illustrated earlier,
appears for $2.4375 < B < 2.6$, where the system forms the
sliding bubble track phase illustrated in Fig.~\ref{fig:13}(b)
at $B = 2.5$.
Here, pinned particles remain trapped in stripes that form tracks
parallel to the driving direction, and the bubbles travel along these
pinned tracks.
For $B \geq 2.6$, the system forms a moving bubble lattice,
as shown in Fig.~\ref{fig:13}(c) at $B = 2.75$,
where it can be seen that when the bubbles
form, they do not fit into the substrate troughs.
The transition to the moving bubble
lattice occurs at the large jump up in $\langle V\rangle$
in Fig.~\ref{fig:12}.
As $F_p$ is varied, the same phases appear for shifted values of $B$.

\begin{figure}
\includegraphics[width=\columnwidth]{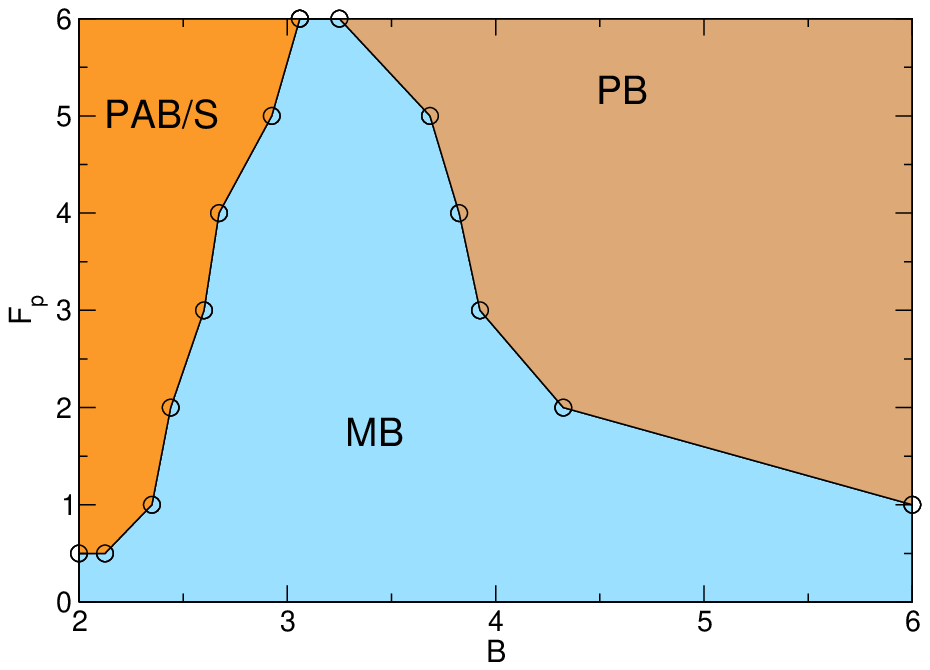}
\caption{Dynamic phase diagram as a function of $F_p$ vs $B$
for the system in Fig.~\ref{fig:12} with $F_D=0.5$, $\rho=0.44$,
and $a_p=2.11$
showing the pinned anisotropic bubble/stripe phase (PAB/S),
the moving bubble lattice (MB), and the
pinned bubble lattice (PB) that appears for higher $B$.}
\label{fig:14}
\end{figure}

In general, the bubbles are less strongly pinned for the
smaller substrate lattice spacing because the size of the bubbles is
larger than the width of the substrate troughs, causing the bubbles
to sit partially
on the potential maxima separating adjacent pinning troughs.
As $B$ increases,
the bubbles shrink, and the velocity goes to zero when
$B$ becomes large enough to permit each bubble to fit entirely within
a single substrate trough,
as shown in Fig.~\ref{fig:13}(d) for $B = 4.5$.
When the driving force is fixed to $F_D=0.5$, the moving bubble phase
continues to flow as the pinning force increases up until
$F_p = 6.0$, when the bubbles become pinned.
In Fig.~\ref{fig:14}, we construct a dynamic phase diagram as
a function of $F_{p}$ versus $B$ for the system in
Fig.~\ref{fig:12}, highlighting the regime in which the moving
bubble lattice can occur.
For $B< 3.0$, the pinned phase
consists of a strongly anisotropic bubble or stripe arrangement,
while for $B > 3.0$, the system forms a pinned bubble lattice.
We note that right along the boundary between the pinned anisotropic
bubble/stripe state and the moving bubble state,
the system is in the plastic sliding bubble track phase,
which is too narrow to highlight in the diagram.

\begin{figure}
\includegraphics[width=\columnwidth]{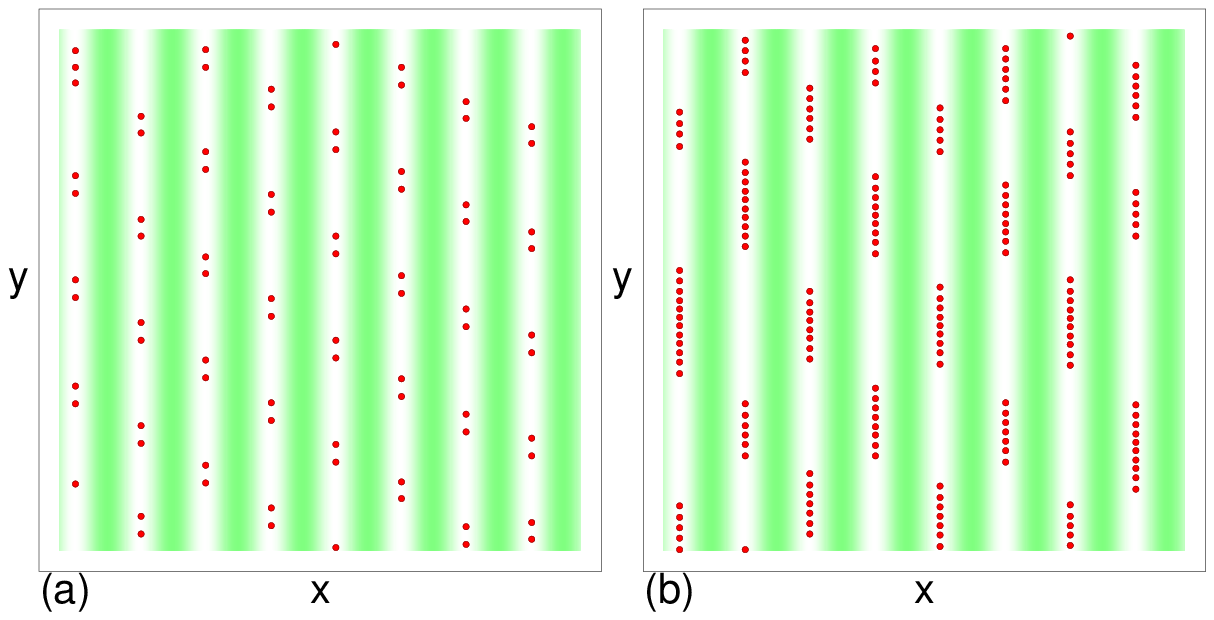}
\caption{The particle positions (red circles) and q1D substrate potential
(green shading) for low particle density systems with
$F_{p} = 4.0$, $a_{p} =4.5$, and $B = 2.85$ at $F_D=0$.
(a) A dimer lattice at $\rho = 0.0617$.
(b) A stripe bubble lattice at $\rho = 0.126$. }
\label{fig:15}
\end{figure}

\begin{figure}
\includegraphics[width=\columnwidth]{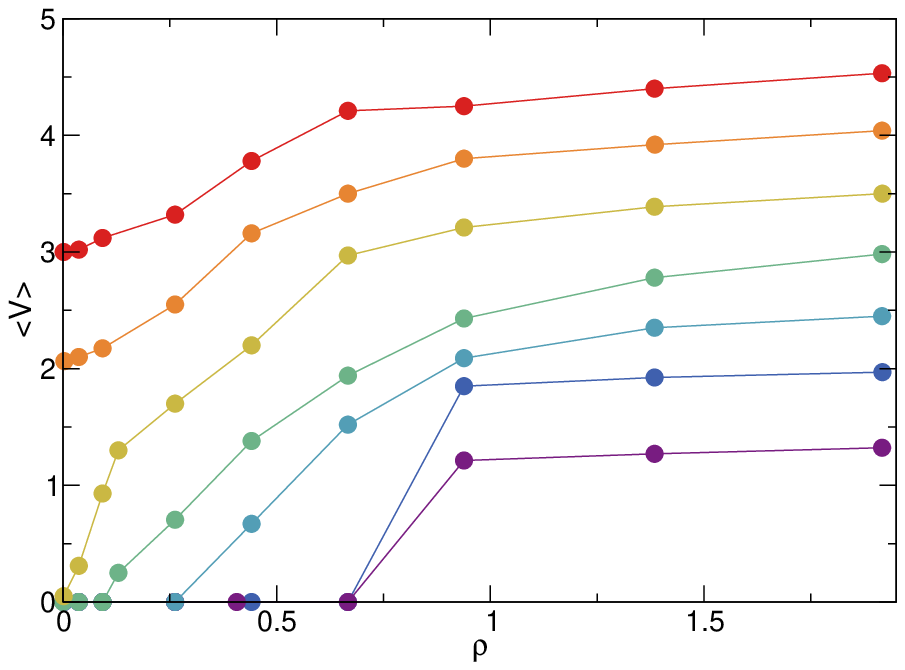}
\caption{$\langle V\rangle$ vs $\rho$ for a system with $F_{p} = 4.0$,
$a_p = 4.5$, and $B = 2.85$ at
$F_{D} = 2.0$, 2.5, 3.0, 3.5, 4.0, 4.5, and $5.0$, from bottom to top,
showing
a general increase in the velocity with
increasing $\rho$ as the bubbles grow in size.
The system is more strongly pinned for the lower densities,
where 1D anisotropic structures appear.}
\label{fig:16}
\end{figure}

To further demonstrate that the larger bubbles are less well pinned, we
fix $B=2.85$ while increasing the particle density in order
to produce large bubbles similar to those shown in
Fig.~\ref{fig:1}(d).
In general, we find that smaller particle densities result in
higher effective pinning.
In Fig.~\ref{fig:15}(a,b)
we show the bubble configurations for
a system with $a_p=4.5$ and $F_{p} = 4.0$ at lower particle densities,
where we find a dimer bubble state for
$\rho=0.0617$ and
a lattice-like arrangement of 1D stripes
at $\rho = 0.126$.
In general, the 1D-like structures are more strongly pinned than
the 2D bubble phases
that form at higher particle densities.
In Fig.~\ref{fig:16} 
we plot $\langle V\rangle$ versus $\rho$ for a system with $F_{p} = 4.0$,
$a_p = 4.5$, and $B = 2.85$
at $F_{D} = 2.0$, 2.5, 3.0, 3.5, 4.0, 4.5, and $5.0$.
When $F_{D} < 4.0$ and $\rho<0.75$, the system is a pinned
1D bubble state similar to those illustrated
in Fig.~\ref{fig:15}.
In general, the average particle velocity increases with
increasing $\rho$ as the bubbles become larger.

\section{Discussion}

There are many further directions to study in this system. For example,
the plastic flow we observe
consists of bubble-to-bubble hopping, so the bubbles
have to break apart; however, there could also be plastic flow
states in which individual
bubbles remain intact, but some bubbles remain pinned while others move,
similar to the
plastic flow
observed in non-bubble particle-like systems.
We expect that this type of bubble lattice plasticity
could appear in systems where the pinning substrate is more heterogeneous.
We have only considered external driving applied parallel to the substrate
modulation direction or $x$ axis,
but the drive could be applied at angles with respect to the $x$ axis,
which would likely produce
a combination of
sliding along the $y$ direction and hopping in the $x$ direction.
The depinning forces could also exhibit commensuration effects based on
how well the bubble size matches the substrate lattice spacing,
so that bubbles whose radii are integer multiples of the substrate
lattice constant
are more strongly pinned.
We have only considered
dc driving, but if the dc drive were combined with ac driving,
we would expect additional Shapiro step-like phenomena
to arise, especially in the sliding bubble lattice state
where there is a strong washboard
frequency.
If thermal effects were included,
it is likely that
the intra-bubble hopping phase would develop
an extended range of creep-like behavior,
and all of the depinning thresholds would shift to
lower values.
In this case,
it would be interesting to compare intra-bubble creep with bubble lattice
creep.

\section{Summary} 

We have investigated a pattern-forming system of particles with competing
long-range repulsion and short-range attraction
driven over quasi-one-dimensional periodic substrates,
and have focused on the bubble regime.
For a fixed driving force, stripe states
are strongly pinned because they can align with the pinning troughs.
In the bubble phase, there can be elastic depinning,
where the moving bubbles maintain their original complement of particles 
and the differential velocity force curves show only a single
peak, or plastic depinning, where individual particles hop from one pinned
bubble to an adjacent pinned bubble.
In the plastic depinning regime, the differential velocity force curves
exhibit two peaks, with the second peak corresponding to the drive at
which all of the bubbles begin to flow but the motion remains plastic
since individual bubbles continuously shed and reabsorb particles.
At higher drives, the system can dynamically reorder into a moving
bubble lattice where no particle shedding occurs.
When the pinning substrate lattice constant is reduced,
we also find a plastic sliding bubble track phase
in which a portion of the particles form tracks consisting of
pinned stripes oriented with the driving direction,
and the remaining particles form bubbles that move over the tracks.
The effectiveness of the pinning, which is visible in both the
depinning threshold and the velocity of the moving bubbles,
depends on the size and flexibility of the
individual bubbles.
When the attractive interaction term is weak and the bubbles are
highly flexible, the bubbles
can distort anisotropically in order to fit between the pinning troughs
and become better pinned.
Increasing the attractive interaction term stiffens the bubbles,
making them more round in shape, but also shrinks their radius.
The stiffer bubbles cannot accommodate themselves to the shape of the
pinning troughs, but once the bubbles drop below the width of the
pinning troughs, they can be well pinned by the substrate.
This leads to a non-monotonic
dependence of the velocity on the strength
of the attractive interaction term,
where there is a pinned state for flexible 
bubbles that can distort into stripe-like shapes, as well
as a reentrant pinned state
that appears when the size of an individual bubble decreases enough that it
can fit inside a pinning trough.
Our results should be relevant to a variety of bubble-forming systems,
including electron bubbles, colloidal particles, and magnetic
skyrmions.

\smallskip

\begin{acknowledgments}
We gratefully acknowledge the support of the U.S. Department of
Energy through the LANL/LDRD program for this work.
This work was supported by the US Department of Energy through
the Los Alamos National Laboratory.  Los Alamos National Laboratory is
operated by Triad National Security, LLC, for the National Nuclear Security
Administration of the U. S. Department of Energy (Contract No. 892333218NCA000001).
\end{acknowledgments}

\bibliography{mybib}

\end{document}